%% file: main.tex
\documentclass[submission]{eptcs}
\providecommand{\event}{GandALF 2010}
\usepackage{breakurl, graphicx, verbatim, enumerate, latexsym}            

\title{A logic for networks}

\author{Massimo Franceschet
\institute{Department of Mathematics and Computer Science, University of Udine \\
           Via delle Scienze 206, 33100 Udine, Italy \\
           Phone: +39 0432 558754 / Fax: +39 0432 558499 \\
           \email{massimo.franceschet@dimi.uniud.it}
           }
        }

\def\titlerunning{A logic for networks}
\def\authorrunning{Massimo Franceschet}

\begin{document}

\maketitle

\begin{abstract}
Networks are pervasive in the real world. Nature, society, economy, and technology are supported by ostensibly different networks that in fact share an amazing number of interesting structural properties. Network thinking exploded in the last decade, boosted by the availability of large databases on the topology of various real networks, mainly the Web and biological networks, and converged to the new discipline of \textit{network analysis} -- the holistic analysis of complex systems through the study of the network that wires their components. Physicists mainly drove the investigation, studying the structure and function of networks using methods and tools of statistical mechanics. Here, we give an alternative perspective on network analysis, proposing a logic for specifying general properties of networks and a modular algorithm for checking these properties. The logic borrows from two intertwined computing fields: XML databases and model checking.
\end{abstract}

\input{introduction}

\input{network}

\input{logic}
\input{conclusion}

\bibliographystyle{eptcs}

\end{document}

%% file: introduction.tex
\section{Introduction} \label{intro}

``\textit{Networks are present everywhere. All we need is an eye for them}'', says Albert-L\'{a}szl\'{o} Barab\'{a}si, in the introduction of his captivating, playful, and elegantly written book about network science \cite{B03}. He is not far from truth. Networks are fundamental tools for modelling and understanding  social, linguistic, biological, technological, and economic complex systems. A \textit{complex system} is made up of a large number of components, or agents, interacting in such a way that their collective behaviour in not a simple combination of their individual behaviour \cite{N03}. Craig Reynolds, an artificial life and computer graphics expert, expressed it as ``\textit{A flock is not a big bird, but the sum of the birds plus the interactions between the birds''} \cite{R87}. 

For decades, we assumed that the components of such complex systems are randomly wired together. In the last ten years, thanks to the wide availability of large databases on the topology of various real networks, many researchers independently showed that such an assumption is wrong: real networks have similar architectures, regardless of their age, function, and scope, that elude the random world \cite{B09}. This provoked the fast growth of the new research field of \textit{network analysis}, the holistic study of structural properties of real networks. This scientific revolution was driven mainly by physicists, because the methods and tools of statistical mechanics are particularly well suited to analyse the patterns of interactions in networks \cite{R06}. Pioneers in this endeavor were Albert-L\'{a}szl\'{o} Barab\'{a}si and Mark Newman \cite{AB02,N03}.
Network analysis addresses questions at three levels of granularity \cite{BE04}: \textit{element-level analysis}, where methods to identify the most important nodes of the network are investigated,  \textit{group-level analysis}, that involves methods for defining and finding cohesive groups of nodes in the network, and  \textit{network-level analysis}, that focuses on topological properties of networks as a whole as well as on theoretical models explaining the generation of empirical networks with certain properties.  

In this work, we throw logic in the arena of network analysis. \textit{Kripke structures} -- networks in which nodes are labelled with a set of properties that hold at the node -- have, in fact, a long tradition in logic: they are the models for the interpretation of modal and temporal formulas \cite{Kr63}. Since the behaviour of a nondeterministic finite state machine can be modelled as a Kripke structure, \textit{modal and temporal logics} are extensively used for the formal specification of properties of hardware and software systems \cite{E90}, and many algorithms and heuristics have being developed to automatically check these properties against the modelled behaviour \cite{CGP99}. Here, we devise a combined logic for the specification of meaningful properties of real networks. The proposed logic combines \textit{XML Path Language} (XPath) properties that look inside the local nodes of the network with \textit{Computation Tree Logic} (CTL) statements that browse the topology of the network. XPath, a simple and elegant node-retrieval language for \textit{Extensible Markup Language} (XML) documents, is one of the most successful technologies for XML data management \cite{XPATH99}. CTL is a blockbuster among logics for formal verification of computer systems \cite{CES86}, effectively applied in the formal verification of safety- and security-critical systems \cite{MWC10}. We furthermore discuss an approach to build a modular model checker that verifies the properties specified in the proposed logic. The model checker exploits an XPath query processor and a CTL model checker. Efficient and scalable implementations exist for both such modules.

The outline of the paper is as follows. In Section \ref{examples} we give some remarkable examples of networks and briefly explain why they deserve attention. Section \ref{analysis} briefly reviews the main contributions to network analysis. Section \ref{logic} describes the combined logic and the corresponding model checker we propose in this paper, while Section \ref{conc} discusses future directions.

%% file: network.tex
\section{Network examples} \label{examples}

Mathematically, a network is a \textit{graph} $G = (V,E)$, where $V$ is a set of \textit{nodes} and $E$ is a set of pairs of nodes called \textit{edges}. In a \textit{directed} network, edges have a direction: an edge is a pair $(x,y)$ where $x$ is the \textit{predecessor} of $y$ and $y$ is the \textit{successor} of $x$. If the network is \textit{undirected}, an edge is a binary set $\{x,y\}$ where the order of the nodes does not matter. We say that $x$ and $y$ are \textit{neighbours}. In a \textit{weighted} network edges are labelled with a numerical weight. In the language of network analysis, nodes with an extraordinary number of edges are called \textit{hubs}.

So, what is the difference between graphs and networks? While a graph is an abstract mathematical object, a network is a concrete, real, and live entity with specific properties. Hence, the investigation of meaningful properties of networks must be driven by the function of the network in the real world and, conversely, discovered properties of networks must be interpreted in terms of such function.  Let us give some remarkable examples of networks and briefly explain why they deserve attention:

\medskip \noindent \textit{The World Wide Web}. This is a directed network in which nodes represent Web pages and edges are the hyperlinks between pages. More precisely, there exists an edge from page $p$ to page $q$ if page $p$ contains at least one hyperlink pointing to page $q$. Usually, the actual number of hyperlinks from $p$ page $q$ is not important and hence the network modelling the Web is unweighted.
Studying the Web as a network is of crucial importance in the field of Web information retrieval. Web search engines, for instance, heavily exploit the Web topology in order to rank Web pages that are returned to the user that issued a query. The PageRank method, which is a major ingredient of Google search engine, is a fitting example \cite{BP98}.

\medskip \noindent \textit{The Internet}. This is a collection of routers linked by various physical lines. The Internet is a growing network with no central control authority. When adding a new node to the Internet, two factors mainly determine the router node to connect to: distance and bandwidth. While distance puts obvious constraints, bandwidth, a measure of connection speed of the router, is typically the dominant factor. This explains the emergence of hubs in the Internet  \cite{FFF99}. The study of Internet topology is crucial to investigate the robustness of the network under failures, which involve nodes randomly, and attacks, which purposely decimate network hubs.

\medskip \noindent \textit{Powerline and airline networks}. These are human-made networks that might be involved in random failures, possibly with cascading effects, as well as targeted attacks \cite{ASBS00,A00}. Clearly, such events on these networks might have catastrophic consequences. The topology of the network directly influences the magnitude and reach of such events. If the network is highly connected and dominated by few hubs, then random failures are generally not problematic, but  attacks aimed to destroy the vital hubs might have Draconian effects.

\medskip \noindent \textit{Citation networks}. An article citation network links scholarly papers through bibliographic references contained in the bibliography of the papers. This network is directed and follows the temporal ordering of papers: we cite the past, not the future. Hence, cycles are very rare, and a citation network closely resembles a directed acyclic graph. However, papers may be aggregated at different levels, forming publication units representing scholars, journals, research fields, regions, nations and even continents. All these bibliometric units can play the role of nodes is a citation network, with edges representing the citations among them. For instance, in a journal citation network, nodes are academic journals, and there is an edge from journal $i$ to journal $j$ if some article published in $i$ cites some article appearing in $j$. Usually, such a network is weighted, with the weight of an edge representing the number of citations between the journals participating in the edge. Citation networks are fundamental tools in \textit{bibliometrics}, the discipline that concerns itself with the study of the dissemination of knowledge through academic publication. In particular, bibliometric indicators like the PageRank-inspired Eigenfactor take full advantage of the topology of journal citation networks \cite{Eigenfactor07}. Citation networks arise also in different contexts like patents and corresponding citations \cite{N94}, published opinions of judges and their citations within and across opinion circuits \cite{LLS98}, and even sections of the Bible and the biblical citations they receive in religious texts \cite{MT05}.

\medskip \noindent \textit{Language networks}. In these networks the nodes are words and the links represent significant co-occurrence in texts, or semantic relationships like synonyms and antonyms. The features of such networks might reflect the evolutionary and social history of lexicons. Furthermore, language disorders like agrammatism, a kind of aphasia in which speech is non-fluent, laboured, halting and lacking in function words, which notably are the most connected words of the network, can be explained in terms of the language network architecture \cite{CS01}.

\medskip \noindent \textit{Conceptual design diagrams}. These diagrams are used to model the concepts and the relationships among concepts of a universe of discourse. The most popular conceptual data model is the Entity-Relationship model, which represents the reality as a diagram in which nodes are entities (like scholar and university) interlinked by various types of relationships (like affiliation) \cite{C76}. Conceptual design is a fundamental step in the design of any database. The identification of central nodes in a conceptual diagram might be important to spot which entities will be most frequently queried by the database users, and hence to optimize the performance of the designed database management system.

\medskip \noindent \textit{Food webs}.  These are networks created by nature. In food webs, species are connected by links telling which species feeds on which other species. The links of these networks seldom go both ways, and hence food webs are also an example of directed networks. Studying food webs is important to understand the ecosystem dynamics. For instance, ecologists believe that hubs of food webs are the keystone species of the ecosystem, paramount in maintaining the stability of the ecosystem. The ecosystem can easily survive if random species are deleted; if, however, hub species are removed, the ecosystem dramatically collapses \cite{SM01,AP09}.

\medskip \noindent \textit{Economic networks}. Market can be viewed as a huge directed multi-relational network. Companies, firms, financial institutions, governments play the role of nodes. Links symbolize different interactions between them, for instance purchases and sales, and the weight of the links captures the value of the transaction. Viewing the economy as a network of interacting actors is useful to make sense of global financial meltdowns, which are provoked by a sequence of failures cascading over the highly connected and interdependent network economy \cite{K97}.
Notably, weighted networks in disguise have been exploited by Wassily W. Leontief to show the input-output dependency relationships among all industries in the economy and the resulting input-output model has been used to estimate the impact on the entire economy of the change in demand (including failure) in any sectors of the economy \cite{L41}. In 1973, Leontief earned the Nobel Price in Economics for his work on input-output tables. The method devised by Leontief has been recently recognised as an early predecessor of the Google PageRank algorithm \cite{F10-pagerank}. 

\medskip \noindent \textit{Metabolic and protein networks}. The nodes of metabolic networks are simple chemicals like water or complex molecules like ATP. The links are the biochemical reactions that take place between these molecules. Moreover, proteins can be viewed as nodes of a complex network in which two proteins are connected if they can physically attach to each other. The robustness of such life maps under failures determines our ability to survive various diseases, and the identification of hub molecules and proteins allow researchers to design effective drugs to cure them \cite{JTAOB00,JMBO01}.

\medskip \noindent \textit{Social networks}. Last but by no means least, social networks link people according to various social relationships, like acquaintance, friendship, collaboration, and sexual relation. They are of paramount importance to understand and anticipate the spread of ideas, innovations, fads, as well as biological and computer viruses \cite{PV01}. For instance, the dominant position of hubs in sexual networks -- people with an extraordinary number of sexual partners -- has been adopted as an explanation of the partially unexpected diffusion and persistence of AIDS epidemic \cite{B03}. Indeed, due to their high connectivity, hubs are easy to be infected and, once infected, they potentially can pass the virus to all linked people. This defies the predictions of classical epidemic models, which are based on the homogeneous, random network hypothesis.
Furthermore, since the seminal works of John R.~Seeley, Leo Katz and Charles H. Hubbell, social networks has been extensively used to measure the social standing of people participating in the network \cite{S49, K53,H65}. The interpersonal directed links in a social network are interpreted as input-output channels for the transmission of influence, and the possibly negative weight of links captures the endorsement strength between individuals.

\section{Network analysis} \label{analysis}

K\"{o}ningsberg was a thriving city close to St.\ Petersburg in eastern Prussia on the banks of the Pregel. The city had seven bridges across the river, and its peaceful citizens wondered if one could ever walk across all seven bridges and never cross the same one twice.  In 1736, Leonhard Euler modelled the K\"{o}ningsberg bridges problem as a graph and mathematically showed that it has no solution. Indeed, nodes with an odd degree must be either the start or the end of an Eulerian path that traverses all bridges once. Hence, an Eulerian path cannot exist if the graph has more than two nodes with odd degree, which was the case for the K\"{o}ningsberg bridges network. Euler is credited as the father of graph theory and the K\"{o}ningsberg bridges network as the first example of a graph \cite{BLW86}. Besides solving the riddle, Euler unintentionally showed that networks have structural properties that limit or enhance their usability.

\textit{Network analysis} is the study of topological properties of real networks \cite{BE04}. It can be performed at three aggregation levels:

\begin{itemize}
\item \textit{Element-level analysis}. At this level, fundamental questions are: ``Which is the most important node of the network?'', or, more specifically, ``How important is a node?'' Importance here is not an intrinsic and permanent feature of the node but, instead, it is an extrinsic and fleeting property that depends on the interactions of the node with the other nodes in the network. Centrality measures include degree centrality, eigenvector centrality, closeness centrality as well as betweenness centrality \cite{F79,N08}. 

\item \textit{Group-level analysis}. This investigation involves methods for defining and finding cohesive groups of nodes in the network. A typical group-level analysis is clustering \cite{A73,JD88}. 
This is an organization process with the goal to put similar objects together. The first step to group similar nodes is to define a similarity function between nodes. When a similarity notion has been defined, we want to organize objects into groups whose members are similar in the defined way and are dissimilar to objects belonging to other clusters.

\item \textit{Network-level analysis}. The focus of this analysis is on properties of networks as a whole as well as on theoretical models explaining the generation of networks with certain properties. Below, we will review typical topological properties of real networks. As for generative models, we mention the random model, also known as Erd\H{o}s-R\'{e}nyi model \cite{SR51,ER60}, Watts and Strogatz model \cite{WS98}, as well as the model of cumulative advantage \cite{S55,P76}, which was later rediscovered and further investigated under the name preferential attachment \cite{BA99,BAJ99}.
\end{itemize}

Despite they emerge in very different contexts, many real networks share remarkable interesting properties. Probably, the most popular feature is the \textit{small world effect}. It is found that in many real webs the mean geodesic distance between node pairs is remarkably short compared to the size of the network. The phenomenon was first told in 1929 by Hungarian poet and writer Frigyes Karinthy in a story entitled ``Chains'' \cite{B03}, and studied in 1967 by Harvard psychologist Stanley Milgram \cite{Mi67}. In a much celebrated experiment Milgram asked American people to get a postcard message to a specified target person elsewhere in the country by sending it directly to the target if the sender knew the target person on a personal basis or otherwise to a personal acquaintance who is most likely to know the target person. Milgram remarkably found that, on average, the length of the chain from the source to the target people was only six. The finding has been immortalized in popular culture in the phrase ``six degrees of separation''.\footnote{``Six degrees of separation'' is the title of a 1990 play written by John Guare and of the following movie directed by Fred Schepisi that was adapted from the play. Both were inspired by the small world phenomenon. The enchanting Alejandro Gonz\'{a}lez I\~{n}\'{a}rritu Oscar-winning film \textit{Babel} is based on the same concept: the lives of all of the characters are dramatically intertwined, although they do not know each other and live in different continents.}

Another interesting property shared by many real networks is \textit{clustering}, also known as \textit{transitivity}. In common parlance, this means that ``the friend of my friend is also my friend''. Given a connected triple $x$, $y$ and $z$ such that $x$ is linked to both $y$ and $z$, an interesting question is: what is the probability that $y$ and $z$ are also connected forming a triangle? In particular, is this probability higher than for randomly chosen nodes? In most real-world networks, the probability of transitively closing a connected triple into a triangle is significantly high. The \textit{clustering coefficient} of a network measures this probability and can be defined as $$C = 3 \frac{N_\bigtriangleup}{N_\wedge}$$ where $N_\bigtriangleup$ is the number of triangles and $N_\wedge$ is the number of connected triples in the network. The factor 3 constrains the coefficient to lie in the range between 0 and 1. For instance, in a much influential sociology paper, \textit{The Strength of Weak Ties}, Mark Granovetter analysed the  social network of people acquaintances and found that our society is far from a random world \cite{G73}. It is structured into strongly connected clusters with few external weak ties connecting these cliques. These weak ties are in fact very strong: they play a crucial role in our ability to communicate outside of our close-knit circle of friends.

Most real instances of networks exhibit a \textit{giant component} and many much smaller components. This is a large group of connected nodes of the graph occupying a sizable fraction of the whole network. The size of the giant component is propositional to the size of the graph. By contrast, the other connected components of the graph are much smaller, usually of logarithmic size with respect to the network largeness. The \textit{diameter} of the network is defined as the largest geodesic distance between pairs of nodes belonging to the giant component.

Finally, real networks possess a node degree distribution with a heavy tail that resembles a \textit{power law}. This means that the probability $P(k)$ that a node has $k$ neighbours is roughly $k^{-\alpha}$, where $\alpha$ is a parameter that usually lies between 2 and 3. It follows that for most real webs the \textit{Pareto principle} applies to node degrees: the overwhelming majority of nodes have low degrees while a few hubs possess an extraordinary number of neighbours \cite{P97}. For instance, the majority of papers published in a journal collect a low share of citations, while there are few citational blockbusters that harvest the majority of citations.
The corresponding networks are called \textit{scale-free}, meaning that there exists no node with a degree that is characteristic for all nodes in the network \cite{B03}. Putting it another way, the degree distribution is highly \textit{skewed} to the right and hence the distribution mean is not a good indicator of central tendency as it is for normal bell-shaped distributions. Notably, scale-free networks are robust under failures but fragile under attacks \cite{AJB00}. A scale-free network is glued together by hubs. Hubs are rare compared to ordinary nodes. Since failures affect random nodes in a network, the probability that a hub is damaged by a failure is relatively low. The price for this resilience, however, is the vulnerability to malicious attacks. These aim to dismantle the hubs of the web, not the ordinary nodes. An attack to few of this hubs is in fact sufficient to break the network into disconnected small pieces.

%% file: logic.tex
\section{A logic for networks} \label{logic}

In this section we propose a logic to specify general properties of real-world networks and we describe computer algorithms to efficiently verify such properties. The logic we propose is inspired by the following observations:

\begin{itemize}

\item In real networks, nodes are real complex objects. Examples include a Web page, an Internet router, an academic publication, an academic journal, a scholar, a biological species, a molecule, a protein, a person (see Section \ref{examples}). These objects are rich of valuable information and meaningful \textit{node properties} can be specified and checked on them. A node property is like a question that can be either true or false at each node object. Importantly, the truth of the property depends only on the information stored inside the node object, and not on the relationships that the node has with other nodes in the network. Examples of such properties are: does the Web page contain the expression \textit{power law} in bold within a figure caption? Is the publication a conference paper published before year 2000? In the journal edited by Moshe Vardi? Does the scholar currently work at MIT? Does the species' name contain the word \textit{Canis}? Is the protein an enzyme? Is the administrator a young CEO?

\item In a network, nodes are connected to each other forming the architecture of the web. \textit{Path properties} are statements that hold true over nodes and involve the relationships that the node has with the other nodes in the network. Hence, while node properties look inside the node objects, path properties investigate the neighbourhood of nodes, and more generally the whole network topology. Examples of path properties are: does the Web page links to Google page? In the author cited by Moshe Vardi? Does that paper cite only papers about network analysis? Can the molecule reach ATP molecule in a chemical path of at most 3 reactions? Is there an arbitrarily long sexual path between the person and Gaetan Dugas? Is there a collaboration path between the scholar and Paul Erd\H{o}s such that all intermediate scholars published more than 100 papers?

\end{itemize}

The ultimate goal of the proposed logic is to integrate the information on the connections of each node with the information about its characteristics. Interestingly, an information-theoretic approach to integrate node and topological information of a network has been recently proposed in \cite{BPM09}.

\subsection{Node properties} \label{node}

We argue that the information contained in a node object is often \textit{semi-structured}, \textit{hierarchical}, and \textit{hybrid}. Semistructured data has a loose structure or schema: a core of attributes are shared by all objects associated to a semistructured schema, but many individual variants are possible. For instance, consider a bibliography containing references to academic publications. All references have in common a small core of attributes, like authors' names, title, and publication year. Different reference types, however, have many specific attributes. For instance: a book has  publisher, a journal article has volume and number, a conference article has proceedings' title and  conference address, a thesis has hosting university. Moreover, some of these attributes might be structured in different ways. For example, we might specify the author name as a unique name or as an arbitrarily long list of names, including space for possibly multiple first, middle, and last name parts.

\textit{Hierarchical} data is composed of atomic elements and structured elements. Atomic elements have a simple flat content. By contrast, structured elements contain nested sub-elements, either atomic or structured. There is no limit to the nesting level of information. The resulting structure is a hierarchy of information, possibly representable as a tree of objects. For instance, a Web page written in XHTML has a hierarchical structure in which tag elements might contain other tag elements. As another example, consider the biological taxonomic hierarchy in which a species, the most basic rank, nests inside a genus, which in turn nests inside a family, and so on.

The term \textit{hybrid} refers to the fact that information mixes both data and text alike. For instance, a bibliographic reference might contain records such as author, title, year, as well as textual information like an abstract. Similarly, objects representing people might contain attributes like name and occupation along with possibly long narrative descriptions, like a CV.

We advocate that this kind of information is best represented with \textit{Extensible Markup Language} (XML) \cite{XML98}. XML, a standard of the World Wide Web Consortium since 1998, has the following unique strengths as a data format \cite{HM02}:

\begin{enumerate}[i)]
\item \textit{simple syntax}. XML is a well-defined format whose documents are easy to create, manipulate, parse by computer software, and read by humans equipped with a basic text editor. Moreover, XML is portable across different computer architectures and programming languages;

\item \textit{semistructured data model}. The flexibility of the XML data model allows one to represent unstructured information and data with a loose schema;

\item \textit{support for nesting}. The hierarchical nature of XML permits to represent complex structures naturally;

\item \textit{support for hybrid information}. Both data-centric and text-centric information can be easily represented within the same XML document.

\end{enumerate}

As an example, consider the XML element describing a fictitious bibliographic reference for a conference paper presented in Figure \ref{bibitem}. Notice that the author element has a loose schema, allowing for possible variations in the name structure. Moreover, the abstract contains a textual summary of the paper, possible formatted using HTML tags.

\begin{figure}[t] 
\begin{small}
\begin{verbatim}
<bibitem key="FH05" type="inproceedings">
  <author>
    <first>Massimo</first>
    <last>Franceschet</last>
  </author>
  <author>
    <first>Elliotte</first>
    <middle>Rusty</middle>
    <last>Harold</last>
  </author>
  <title>Modal logic and navigational XPath</title>
  <booktitle>Workshop Methods for Modalities</booktitle>
  <pages>156-172</pages>
  <year>2005</year>
  <abstract>
    Three decades past, the <em>relational</em> empire conquered
    the <em>hierarchical</em> hegemony. Today, an upstart challenges
    the relational empire's dominance, threatening the
    return of hierarchy.
 </abstract>
</bibitem>
\end{verbatim}
\end{small}
\caption{A bibliographic item in XML format.}
\label{bibitem}
\end{figure}

A plethora of XML technologies has been developed since XML has been recommended by W3C in 1998. These allow to define schemas and validate documents against them, query XML documents at different levels of complexity, transform the content of XML documents, and access XML documents using programmatic interfaces. Furthermore, native XML databases have been built to store very large data sets in XML format and to efficiently retrieve fragments of them. In particular, the \textit{XML Path Language} (XPath) \cite{XPATH99} is a simple query language to retrieve elements from XML documents. XPath grabbed a lot of attention from the professional and academic communities alike (see \cite{BK08} for a survey). There are good reasons for such a popularity: XPath represents a fundamental technology in the XML world since it is shared by different important XML technologies, like XSchema (to define XML schemas), XQuery (to query XML databases), and XSLT (to transform XML documents). Furthermore,  XML documents can be represented as trees, a well-known data structure in computer science \cite{CSRL01}, and the navigational core of XPath can be interpreted as a modal logic \cite{BdeRV01}, a language well-digested by computational logicians. It follows that the expressive power and the computational properties of XPath have been extensively investigated within the computing community. 

An XPath query is a location path \texttt{/step$_1$/step$_2$/\ldots /step$_k$}. Each location step in the path has the form \texttt{axis::test[filter]}, where \texttt{axis} is a navigation modality (child, descendant, parent, ancestor, sibling, and more), \texttt{test} is a check on the type of the XML elements (element, attribute, and so on), and \texttt{filter} is an optional Boolean combination of location paths that is used to percolate the set of selected XML elements. XPath queries are evaluated at XML elements of XML documents. Since an XML document can be represented as a tree of nodes, it is convenient to interpret the evaluation of an XPath query over a tree. A location step is evaluated at a context node of the XML tree: it traverses the tree starting from the context node along the modality specified by the axis part, retrieves all encountered nodes of the type specified in the test part, and returns only those nodes that satisfy the filter expression. An XPath location path is evaluated as follows: starting from the root of the XML tree, the nodes reached by the first location step are retrieved and given in input as context nodes to the evaluation of the second step, and so on. The result of the query is the outcome of the last location step.

We propose the use of XPath filters to specify properties of node objects of real networks. Since filters are Boolean combination of location paths, they retain the full expressive power of XPath. Moreover, the evaluation of a filter on a network node element results in a Boolean value (either true or false). For example, recall the above XML element describing a bibliographic item. The following filter specifies the property that the publication has a single author, has been published later than 2007, and contains the word \textit{XML} emphasised in the abstract:

\begin{small}
\begin{verbatim}
count(author) = 1 and (year > 2007) and contains(abstract/em, "XML")
\end{verbatim}
\end{small}

To check the node properties, one can take advantage of a query processor for XML as follows. First, encode each node as an XML element and collect such elements in an XML document like this:

\begin{small}
\begin{verbatim}
<network>
  <node key="k1">...</node>
  <node key="k2">...</node>
  <node key="k3">...</node>
  ...
</network>
\end{verbatim}
\end{small}

Each network node is encoded as a  \texttt{node} XML element and it is identified by a unique key value. Furthermore, write the node property as an XPath filter \texttt{F}. Finally, evaluate the query \texttt{/network/node[F]} over the constructed XML document using a query processor for XML. The result of the query evaluation is the set of nodes that satisfy the specified node property.

XPath queries can be evaluated against an XML document in polynomial time with respect to both query and data complexity \cite{GKP02}. In particular, queries expressed in the navigational fragment of XPath -- the core that contains features to navigate the XML tree only -- can be evaluated in linear time with respect to the length of the query and the size of the document. Many scalable and efficient query processors exist for XML databases, a notable example is BaseX \cite{BaseX07}.

\subsection{Path properties} \label{path}

Node properties are the building blocks of path properties. Path properties are statements that talk of the network topology -- how each node is related to the other nodes -- and are defined in terms of local node properties. For instance, we might specify a path property that holds true over those nodes of the network that satisfy a certain node property and that can reach through a path of edges a node that verifies some other node property.

Since the pioneering work of Amir Pnueli \cite{P77}, modal and temporal logics have been extensively used to specify properties of computer systems \cite{E90}. A \textit{modal logic} is a formal language build up from atomic propositions, Boolean connectives, and modal operators, called \textit{modalities} \cite{BdeRV01}. When the modalities have a temporal connotation,  modal languages are referred to as \textit{temporal logics}. Modal and temporal logics are interpreted over \textit{Kripke structure}, which are graphs whose nodes are labelled with a set of properties that hold at the corresponding node \cite{Kr63}. Since the computational behaviour of a nondeterministic finite state machine can be modelled as a Kripke structure, in which nodes are computation states and edges symbolize the evolving computation from state to state, modal and temporal logics can be used as specification languages for computer systems. A \textit{model checker} is a software that, given a Kripke structure and a formula, retrieves the set of nodes of the structure that satisfy the formula \cite{CGP99}. Furthermore, if a node fails to verify the formula, the model checker produces a counterexample showing how the property can be falsified.

We propose to use the language of the branching temporal logic \textit{Computation Tree Logic} (CTL) to specify path properties over networks \cite{CES86}. This is a simple and efficient logic, widely exploited for computer system verification. Notably, the inventors, Edmund Melson Clarke, Ernest Allen Emerson, and Joseph Sifakis, won the 2007 \textit{Turing Award} (the Nobel Prize of computing) for their pioneering work on program verification. The language of CTL is build up from atomic propositions, Boolean connectors $\wedge$ for \textit{and}, $\vee$ for \textit{or}, and $\neg$ for \textit{not}, and path quantifiers. Path quantifiers are defined as follows ($\varphi$ and $\psi$ are CTL formulas):

\begin{itemize}
\item $\mathbf{EX} \varphi$ means that there exists at least one neighbour of the current node where $\varphi$ holds;
\item $\mathbf{AX} \varphi$ means that for all neighbours of the current node $\varphi$ holds;
\item $\mathbf{EF} \varphi$ means that there exists at least one node reachable trough a path from the current node where $\varphi$ holds;
\item $\mathbf{AF} \varphi$ means that for all paths from the current node $\varphi$ holds at some node;
\item $\mathbf{EG} \varphi$ means that there exists at least one path from the current node where $\varphi$ always holds;
\item $\mathbf{AG} \varphi$ means that for all nodes reachable trough a path from the current node $\varphi$ holds;
\item $\mathbf{EU}(\varphi, \psi)$ means that there exists at least one path from the current node on which $\varphi$ holds until at some node $\psi$ holds;
\item $\mathbf{AU}(\varphi, \psi)$ means that for all paths from the current node $\varphi$ holds until at some node $\psi$ holds.
\end{itemize}

If the network is directed, it is useful to add also the inverse path quantifiers $\mathbf{\overline{E}}$ and its dual $\mathbf{\overline{A}}$. While ordinary path quantifiers follow the direction of edges, inverse path quantifiers navigate along the inverse direction. For instance, on a directed graph, $\mathbf{EX} \varphi$ means that there exists at least one node pointed to by the current node where $\varphi$ holds, while $\mathbf{\overline{E}X} \varphi$ means that there exists at least one node that points to  the current node where $\varphi$ holds.

The \textit{model checking problem} for CTL -- the problem of checking which states of a given graph satisfy a given CTL formula -- has been thoroughly investigated in computer science \cite{CES86,CGP99}. Model checking for CTL can be solved in linear time $O(k \, (n+m))$, where $k$ is the query complexity, $n$ and $m$ are the number of nodes and edges of the graph, respectively. Efficient model checkers for CTL are available, e.g., NuSMV \cite{CCGR99}.

\subsection{An XML path logic for networks} \label{combine}

The logic for networks we propose is named \textit{XML Path Logic} (XPL, for short). It combines XPath node properties with CTL path statements, taking advantage of the expressive power and computational speed of both XPath and CTL languages. Precisely, an XPL formula is defined as a CTL formula (possibly extended with inverse path quantifiers) in which atomic propositions are replaced by XPath filters. This form of logic combination is known as \textit{temporalization} \cite{FG96}.

By way of example, we encode in XPL the path properties given at the beginning of Section \ref{logic}:

\begin{itemize}

\item \textit{Does the Web page links to Google page?} \\
\smallskip
$\mathbf{EX} \texttt{(title = Google)}$

\item \textit{In the author cited by Moshe Vardi?} \\
\smallskip
$\mathbf{\overline{E}X} \texttt{((first = Moshe) and (last = Vardi))}$

\item \textit{Does that paper cite only papers about network analysis?}  \\
\smallskip
$\mathbf{AX} \texttt{contains(keywords, "network analysis"})$

\item \textit{Can the molecule reach ATP molecule in a chemical path of at most 3 reactions?} \\
\smallskip
$\mathbf{EX} \texttt{(name = ATP)} \vee \mathbf{EXEX} \texttt{(name = ATP)} \vee \mathbf{EXEXEX} \texttt{(name = ATP)}$

\item \textit{Is there an arbitrarily long sexual path between the person and Gaetan Dugas?} \\ $\smallskip
\mathbf{EF} \texttt{((first = Gaetan) and (last = Dugas))}$

\item \textit{Is there a collaboration path between the scholar and Paul Erd\H{o}s such that all intermediate scholars published more than 100 papers? } \\
\smallskip
$\mathbf{EU}(\texttt{count(paper) > 100}, \, \texttt{(first = Paul) and (last = Erdos)})$

\end{itemize}

Model checking algorithms for combination of logics have been investigated in \cite{FMdR04}. In the following, we sketch a modular model checker for the combined logic XPL. The model checker uses two modules: an XPath query processor  and a CTL model checker. Let $N$ be a network and $p$ be a path property we want to check on $N$. The model checker works as follows. First, we encode the nodes of the network $N$ in XML as described in Section \ref{node} obtaining an XML document $D$. Moreover, we  translate the property $p$ into a formula $\varphi$ in the language of XPL. Then, we run the following algorithm:

\begin{enumerate}
\item \textit{Query evaluation step}. For each XPath filter \texttt{F} contained in $\varphi$, use the XPath query processor to evaluate the query \texttt{/network/node[F]} over document $D$ retrieving all nodes that satisfy the filter and label such nodes in the network $N$ with the corresponding atomic proposition $p_\texttt{F}$;

\item \textit{Formula replacement step}. Replace in $\varphi$ each XPath filter \texttt{F} with the corresponding atomic proposition $p_\texttt{F}$, obtaining a pure CTL formula $\psi$;

\item \textit{Model checking step}. Model check $\psi$ on the node-labelled network $N$ with the CTL model checker retrieving all nodes of the network that satisfy $\psi$. These are also the nodes that verify the original property $p$.

\end{enumerate}

The complexity of the combined model checkers for XPL depends on the complexity of the used modules. CTL model checking can be performed in combined linear time (the product of the formula length and of the graph size). The complexity of XPath query evaluation depends on the fragment of XPath used to encode the path property. Full XPath queries can be evaluated in polynomial time; in particular the evaluation problem for the navigational fragment of XPath can be solved in combined linear time (the product of the query length and of the document size) \cite{GKP02}. Let $n$ and $m$ be the number of nodes and edges of the network $N$, respectively, and $k$ be the length of the XPL formula $\varphi$. We may assume that each node of the network is encoded in an XML element of constant size, hence the size of the entire XML document $D$ is $O(n)$. If we use navigational XPath to encode node properties, then the query evaluation step  of the algorithm costs $O(k n)$. The formula replacement step costs $O(k)$. Finally, the model checking step costs $O(k (n+m))$. All in all, the complexity of the combined model checker for  XPL is $O(k (n+m))$, hence linear in both the length of the formula and the size of the network. If we use a more expressive fragment of XPath as the node language, then the complexity of the combined model checker is still polynomial but dominated by the cost of query evaluation for the XPath fragment.

%% file: conclusion.tex
\section{Future work} \label{conc}

We proposed a logic perspective for the analysis of real networks. To be sure, we only scratched the surface of the potential of the logic approach to network analysis. At least two important open questions remain:

\begin{enumerate}
\item Can we take advantage of the logic method to study novel meaningful properties shared by real networks?

\item Can we exploit the structural features of real networks, namely small geodesic distance, clustering, and power law degree distribution, to speed up the verification of logic properties?

\end{enumerate}

We think that the answer to these questions will determine the success or the failure of the logic approach to the emerging science of networks.